\documentclass[11pt]{article}
\usepackage{setspace}
\raggedright
\usepackage{verbatim}
\usepackage{amsmath,amsfonts,amsthm,bm}

\usepackage{hyperref}
\usepackage{multirow}
\usepackage{setspace} 
\usepackage[natbibapa]{apacite}

\usepackage[shortlabels]{enumitem}
\usepackage{fullpage}
\usepackage{xcolor}

\usepackage{float}
\usepackage{caption}
\usepackage{subcaption}
\usepackage{booktabs}
\usepackage{dsfont}

\RequirePackage{graphicx,fancyvrb}

 \newcounter{paranum}
 
 \newcommand{\Sec}[1]{\vspace{10pt}\noindent\textbf{ #1}}
  \newcommand{\SmSec}[1]{\vspace{10pt}\noindent\textit{ #1}}

\usepackage{fancybox}
\usepackage{ragged2e}

{\vskip 0em\VerbatimEnvironment
\begin{Sbox}\begin{BVerbatim}}
{\end{BVerbatim}%
\end{Sbox}
\hspace{0.3in} \TheSbox\vskip 0em}
\usepackage[title]{appendix}

\begin{document}
\title{Local Randomization Regression Discontinuity Designs when Test Scores are the Running Variable }

\author{Sophie Litschwartz \\ Abt Associates Inc}

\maketitle 
\Sec{Abstract}
\\
\begin{singlespace} 
\noindent

Explanations of the internal validity of regression discontinuity designs (RDD) generally appeal to the idea that RDDs are ``as good as" random near the treatment cut point. \citet*{cattaneo2015} are the first to take this justification to its full conclusion and propose estimating the RDD local average treatment effect (LATE) the same as one would a randomized experiment. This paper explores the implications of analyzing an RDD as a local random experiment when the running variable is a test score. I derive a formula for the bias in the LATE estimate estimated using the local randomization method, $a\rho\Delta$. Where $a$ is the relationship between latent proficiency and the potential outcome absent treatment, $\rho$ is the test reliability, and $\Delta$ is the distance between the treatment and control running variable value. I use this quantification of the bias to demonstrate that local randomization and related design based methods for estimating RDDs are problematic when the running variable is test score or other human developed measure (e.g., medical tests). 
	\end{singlespace} 
\doublespacing

\noindent {\it Keywords:}  latent variable models; regression discontinuity designs; test scores 

\newpage

\Sec{Introduction}

		Regression discontinuity inference is sometimes explained as follows: two different students sit in two different classrooms taking the same math test. The two different students have the same math proficiency, but outside the window of one student's classroom, two dogs start barking at each other. The student with the dogs outside is distracted, misses a question they would normally get correct, and fails the math exam. The student with no dogs gets the question right and passes the math exam. The student who fails the exam is given extra math support, and the student who passes is not. Since the appearance of the barking dogs is random, the student who passes can be used as a control for the student who fails. 

	There is a problem with this story; on average, students who get the minimum passing test score are more proficient than students who get one point below passing. The barking dogs story looks at two individual students with identical proficiency, whose only difference is the presence of barking dogs. However, on average, students who get the minimum number of questions correct to pass will experience as many barking dogs as the students who fall one question short of passing. Therefore in expectation, students who fail the exam by one question are not a good control for students who pass the exam by one question because, on average, the passing students are one question more proficient than the failing students.  

	In practice, this limitation of the barking dog story is not a problem for most regression discontinuity design (RDD) research. Heuristic explanations of RDD in the barking dog genre are meant to provide the intuition for RDD. They are not intended to be formal articulations of the RDD identification assumption. Instead, most RDD research relies on the continuity assumption first articulated by \citet*{hahn2001}, which is that the potential outcome of interest is continuous through the cut point. This continuity assumption is most commonly operationalized using local linear regression(LLR). Under LLR, the outcome of interest (e.g., passing a math course) is modeled as a linear combination of the running variable (e.g., math test score) and treatment status (e.g., receiving extra math support). The treatment effect is solved for using a traditional OLS estimator but only using points within some optimal bandwidth of the treatment cut point. 
	
	Still, there is a tension in the literature between relying on the looser continuity assumption for identification and a desire to treat RDD identification ``as good as random" near the cutoff. \citet*{lee2010} argues that as long as there is some measurement error in the running variable and individuals cannot precisely control their running variable value, then an RDD is equivalent to a randomized experiment in the region of the cutoff. Papers that provide overviews of RDD methods also frequently explain RDD as a local randomized experiment \citep{imbens2008,lee2010}. There is also empirical evidence that RDDs provide the same estimates of the local average treatment effect as RCTs \citep{chaplin2018}. 

	However, even researchers that appeal to an ``as good as random" articulation of RDD typically analyze their data using the LLR method suggested by \citet{hahn2001}. \citet*{cattaneo2015} were the first to advocate analyzing RDDs as a locally randomized experiment; with further extensions of this framework produced by \citet*{cattaneo2016} and \citet*{cattaneo2017}. In this local randomization framework, as in a fully randomized experiment, the LATE is calculated by comparing the mean outcome of treated observations near the cut point to non-treated observations near the cut-point. This analysis method is appealing because it moves RDD analysis from model based inference to design based inference. It also allows standard errors to be estimated using randomization inference, which may more accurately capture the sample size used to estimate the LATE.  
	
	Local randomization has subsequently become a mainstream part of the RDD methodological toolkit, and local randomization regularly appears in practitioner guides for RDD \citep{skovron2015, abadie2018, cattaneo2019, cattaneo2021, cunningham2021}. In particular, local randomization is often recommended as a method for estimating RDDs with discrete running variables \citep{skovron2015, cattaneo2017choice, cattaneo2021}. In addition, further research has expanded the original local randomization RDD framework. For example, \citet{keele2015} incorporate matching on observed covariates into local randomization analysis, and \citet{mattei2017} provide a formulation of the Stable Unit Treatment Value Assumption applicable to local randomization RDD. 
	
	In this paper, I explore the implications of using the local randomization method for estimating RDD local average treatment effects when the running variable is a test score. Test scores are a common running variable choice in education because educational interventions are frequently assigned based on test scores. More broadly, test scores serve as a framework for examining the larger class of RDDs, where the running variable is a noisy measure of an underlying latent quantity. 

	I show there is bias in the local randomization RDD LATE estimator and that bias is a multiplicative combination of the relationship between the latent construct and the outcome, the test reliability, and the distance between points on the running variable scale. This bias emerges because test scores are discrete, and the RDD LATE can not be estimated at the limit. However, this discreteness problem is not unique to test scores. All empirical data is to some extent discrete because all measurement methods have finite precision. In addition, sample size constraints require the pooling of data across different points on a continuous scale for it to be analyzed as a random experiment. I also show how this bias can escape detection in covariate based placebo tests and that matching on observable characteristics is slow to reduce the estimator's bias. Finally, I discuss the practical complications with incorporating estimates of test reliability into design based estimators of RDD treatment effects. Taken together, this paper provides a basis for assessing the conditions under which local randomization is an appropriate framework for RDD analysis and when this framework will produce unacceptably biased results.  
	
		This work builds upon prior research by \citet*{Sekhon2017}, who use test scores as a stylized example to demonstrate theoretical limitations in interpreting RDDs as local experiments. However, in this paper, I use test scores not as a stylized example but as concrete objects with their own statistical properties, which can be leveraged to quantify the implications of analyzing an RDD using local randomization. In addition, \citet{Sekhon2017} treat test scores as continuous in their model, while this paper contributes to a body of literature focused on how to model RDDs when there is a discrete running variable \cite{Armstrong2020, Calonico2018, Calonico2014, Kolesar2018, lee2008}.      
		
		In addition, this research contributes to a recent body of literature that applies a latent variable model to RDD.  \citet*{Soland2022} use a latent variable model to demonstrate how RDDs can be analyzed using Structural Equation Modeling (SEM).  Similar to other researchers working within the local randomization framework, \citet*{eckles2020} use a latent variable model to create a design based RDD estimator. Specifically, \citet{eckles2020} use reliability estimates of the running variable to derive an RDD estimator based on re-weighting the treatment and control samples.

\Sec{Latent Variable Regression Discontinuity Model}

Under the potential outcomes framework the regression discontinuity model for an individual $i$ can be written as follow:

$$ Y_{0i} = g_0(X_i) + \epsilon_i , \;  \epsilon_{i} \stackrel{iid}{\sim} N(0,\sigma_{\epsilon}^2)$$  
$$ Y_{1i} =  Y_{0i}+g_1(X_i) $$ 
    \[ Y_i= \begin{cases} 
          Y_{0i} & X_i \geq c \\
            Y_{1i} & X_i < c
       \end{cases}
    \]

\noindent where $Y_{0i}$ is the outcome absent treatment, $Y_{1i}$ is the outcome with treatment, $X_i$ is the value of the running variable, and $c$ is cut value that determines treatment. The treatment effect ($Y_{1i} -  Y_{0i}$) for individual $i$ is therefore $g_1(X_i)$.

	When $X$ is a test score, $X$ is a noisy estimate of a true latent proficiency $\theta$. Under a classical test theory model, we can write $X_{i}$ as a linear combination of $\theta$ and a random error term $\omega$ such that:
		
		$$X_{i}=\theta_{i} + \omega_{i}, \; \omega_{i} \stackrel{iid}{\sim} N(0,\sigma_{\omega}^2)$$

	Under a latent variable model framework, $Y$ is not a function of $X$, the noisy estimate, but $\theta$ the underlying latent quantity $X$ measures. Combining the classical test theory model with the potential outcomes RDD model, the RDD problem can now be restated as:
	
$$ Y_{0i} = g_0(\theta_i)  + \epsilon_i $$    
$$ Y_{1i} =  Y_{0i}+g_1(\theta_i) $$ 
    \[ Y_i= \begin{cases} 
          Y_{0i} & X_i \geq c \\
            Y_{1i} & X_i < c
       \end{cases}
    \]
$$ X_i = \theta_i + \omega_{i}$$ 
$$ \epsilon_{i} \stackrel{iid}{\sim} N(0,\sigma_{\epsilon}^2), \;  \omega_{i} \stackrel{iid}{\sim} N(0,\sigma_{\omega}^2), \; Cor(\epsilon_{i}, \omega_{i})=0$$

\Sec{Local Randomization Bias}

	This section uses the latent variable RDD model from the previous section to derive the expected bias in the local randomization treatment effect estimate. In order to do this I make two simplifying assumptions, over the interval $(c,c-\Delta)$:
	\vspace{-1cm}
	\begin{singlespace}
		\begin{center}
	 \begin{enumerate}[leftmargin=3cm]
	 \item  $g_0(\theta_i)$ is linear.
	    \item  $g_1(\theta_i)$ is constant and equals the treatment effect $\tau$.
	 \end{enumerate}
	 \end{center}
	 	\end{singlespace}
		
	Simplifying assumption one should not be considered overly restrictive. It is expected that sufficiently close the treatment cut $g_0(\theta_i)$ can be approximated with a linear function. However, simplifying assumption two is a strong assumption because it assumes a constant treatment effect across values of the running variable. In Appendix \ref{sec:altderivation}, I provide derivations where this second assumption is relaxed, and I arrive at the same bias estimate. Using these assumptions, the simplified RDD model is written as follows:
		
			$$Y_i=a\theta_{i}+\tau \mathds{1}(X_i <c) + \epsilon_i , \; \epsilon_{i} \stackrel{iid}{\sim} N[0,\sigma_y^2]$$
 			$$X_i = \theta_i+ \omega_{i}, \; \omega_{i} \stackrel{iid}{\sim} N[0,\sigma_{\omega}^2]$$		
			
	Local randomization inference treats the RDD as a randomized experiment in the region near the cut point. In the discrete case, I define the control analysis sample as the observations where $X=c$ and the treated analysis sample as the observations where $X=c-\Delta$, where $\Delta$ is the distance between points on the discrete running variable scale. Using a design based difference in means estimator, the treatment effect is estimated as:

	$$  \hat{\tau}= E[Y_1 | X=c-\Delta] - E[Y_0 | X=c]$$
	$$=E[a\theta+\tau+\epsilon | X=c-\Delta]-E[a\theta+\epsilon | X=c]$$
	$$=aE[\theta | X=c-\Delta]+\tau-aE[\theta | X=c]$$
	
According to Kelley's formula \citep{kelley1947}: 	
$$E[\theta | X=x] = \rho x +(1-\rho) E[X] \; , where \; \rho=\frac{\sigma_\theta^2}{\sigma_X^2} $$
	
Therefore the estimator bias can be written as:
		$$  \hat{\tau}-\tau= a( \rho(c-\Delta)+(1-\rho) E[X] - ({\rho}c+(1-\rho)E[X]))$$
		$$  \hat{\tau}-\tau= -a\rho\Delta$$

	The bias is a function of three things: the relation between the potential outcome absent treatment and the latent proficiency ($a$), the test reliability ($\rho$), and the distance between adjacent points at the cutpoint ($\Delta$). If any of these three quantities are zero, then the local randomization bias is zero. In the limit, where $a=0$ or $\rho=0$, we can consider the RDD to be a fully randomized experiment. This might occur if $X$ is not an observed test score but a randomized lottery number. The lottery number has no relationship to either true proficiency ($\theta$) or the potential outcome absent treatment ($Y_0$). The case where $\Delta->0$ is where the data is truly continuous, and the RDD analysis converges to the idealized RDD where the LATE is estimated at the true limit. 

	However, test score RDDs are not situations where this bias will generally be close to zero. First, test scores are strongly correlated with a wide range of outcomes \citep{heckman2006,hanushek2009}, leading to large $a$ values. Second, tests are designed to be reliable, and conventional internal consistency reliability estimates are between .8 and .95 \citep{ho2015}. Finally, test scores typically have a non-trivial value for $\Delta $; for example, both of the 2019 SAT sections had a gap of about .1 standard deviations between score bins \citep{CollegeBoard2019}. Taken together, these facts mean that in most cases, local randomization is not going to be a good method for estimating the RDD LATE in studies that use test scores as a running variable. 

\Sec{Placebo Test}

	  \citet{cattaneo2015} does not ignore the potential for bias in their local randomization analysis framework. The local randomization framework requires there to be a region around the cut point where there is no relationship between the running variable and $Y_0$, the orthogonality assumption. When this assumption holds, $a$ is zero and the bias is also zero. However, the method  \citet{cattaneo2015} give for testing the orthogonality assumption still allows for situations where the RDD LATE is biased.
	  	  
	 \citet{cattaneo2015} propose using a placebo test to determine a region around the cut point where this orthogonality assumption holds. In this test, $Y$ is replaced with a covariate $Z$ (e.g., student race), which is not affected by the treatment. The difference in means between treatment and control is estimated for increasingly large values of $\Delta$. The orthogonality assumption is said to hold for all the $\Delta$ such that $ \hat{\tau}_{Z} = 0$ can not be rejected. The largest value of $\Delta$ when the orthogonality assumption is determined by this test to hold is used as the study bandwidth. 
	 
	 I evaluate this placebo test using a new version of the simplified RDD model where $Z$ is the outcome instead of $Y$. The new model is as follows:  
	$$Z=b\theta_{i}+\tau \mathds{1}(X_i <c) + \epsilon_i , \; \epsilon_{i} \stackrel{iid}{\sim} N[0,\sigma_z^2]$$
\noindent By construction $\tau_Z=0$ and so from the results in the previous section it can be concluded that:
		$$ \hat{\tau}_Z= -b\rho\Delta$$
		
The conditions of the placebo test imply that this $\hat{\tau}_Z$ will be just below the minimum detectable value ($MDV - \delta$ ). $ \hat{\tau}_Z$ will end up close to the $MDV$ because the placebo test procedure for determining the optimal $\Delta$ keeps increasing $\Delta$ until $\hat{\tau}$ is just below the MDV. Solving for $\rho\Delta$ leaves $\rho\Delta=-\frac{MDV - \delta}{b}$. Therefore, the LATE bias will be $-\frac{a}{b}{(MDV-\delta)}$. 

	If the true proficiency ($\theta$) is more predictive of the covariate ($Z$) than the outcome ($Y$), then $a<b$ and the bias will be less than the $MDV$. In this case, when the true LATE is zero, the local randomization estimator will correctly, in most cases, not detect a statistically significant effect even with nonzero bias. However, it is expected that generally $\theta$ will be more predictive of $Y$ than $Z$ and $a>b$. Test scores are specifically designed to be predictive of outcomes of interest to education policymakers, which are also the outcomes frequently measured in education policy research. Tests like the ACT and the SAT partially derive their validity from their ability to predict the GPAs of first-year college students. On the other hand, test makers try to minimize the amount to which they are just capturing immutable characteristics, which are precisely the $Z$ variables commonly used in placebo tests. This dynamic means that most often $a>b$ and the placebo test proposed by \citet{cattaneo2015} will not fully protect against bias in the LATE estimate.

\Sec{Matching}

	One potential method for dealing with the bias in the local randomization method, as proposed by \citet{keele2015}, is to match observations within score buckets. In this section, I will derive a new bias equation that accounts for matching and shows that matching does a poor job of reducing the bias in the LATE. 
	
	Take a new example where $Y$ still depends on $\theta$, and $X$ is a noisy estimate of $\theta$. Now assume $\theta$ is a linear combination of an observed variable $V$ and an unobserved residual $U$. Such that:
	$$Y=a\theta_{i}+\tau \mathds{1}(X_i <c) + \epsilon_i , \; \epsilon_{i} \stackrel{iid}{\sim}  N[0,\sigma_y^2]$$
  	$$X_i = \theta_i+ \omega_{i}, \; \omega_{i} \stackrel{iid}{\sim}  N[0,\sigma_{\omega}^2]$$
	$$\theta_i=dV_{i}+U_{i}$$

\noindent Note that $V$ is measured without error and that $Y$ only depends on $V$ through $\theta$. The new matching estimate for the treatment effect is:
	$$ \hat{\tau}= \sum_{V} {P(V=v)(E[Y_1 | X=c-\Delta,V=v] - E[Y_0 | X=c,V=v])}$$
	$$= \sum_{V} {P(V=v)(a(E[\theta | X=c-\Delta,V=v]-E[\theta | X=c,V=v])+\tau)}$$
	$$E[\theta | X=x,V=v]=\frac{\sigma_{\theta|V=v}^2}{\sigma_{X|V=v}^2}x+\frac{\sigma_{\omega|V=v}^2}{\sigma_{X|V=v}^2}E[X|V=v]$$

\noindent Define $R^2$ to be the percent of the variance of $\theta$ explained by $V$, such that $R^2=\frac{\sigma_{dV}^2}{\sigma_\theta^2}$. Therefore:
	$$Var(\theta | V=v)=Var(dV|V=v)+Var(U|V=v)=Var(U)=(1-R^2)\sigma_\theta^2$$	
		$$Var(X | V=v)=Var(\theta | V=v)+Var(\omega | V=v)=(1-R^2)\sigma_\theta^2+\sigma_\omega^2$$	
	$$E[\theta | X=x,V=v]=\frac{(1-R^2)\sigma_\theta^2}{(1-R^2)\sigma_\theta^2+\sigma_\omega^2}x+\frac{\sigma_{\omega}^2}{(1-R^2)\sigma_\theta^2+\sigma_\omega^2}E[X|V=v]$$
		$$ \hat{\tau}= \sum_{V} {P(V=v)(\tau-a\Delta\frac{(1-R^2)\sigma_\theta^2}{(1-R^2)\sigma_\theta^2+\sigma_\omega^2}})$$

The new equation for the bias is: 
$$ \hat{\tau}-\tau=-a\Delta\frac{(1-R^2)\sigma_\theta^2}{(1-R^2)\sigma_\theta^2+\sigma_\omega^2}$$

 if we assume that $X$ is standardized such that $\sigma_X^2=1$ then the bias reduces to:
 $$- \hat{\tau}-\tau=-\frac{1-R^2}{1-R^2\rho} a\Delta\rho$$

This means that in the matching case, the bias is reduced by a factor of $\frac{1-R^2}{1-R^2\rho}$, which means increasing $R^2$ reduces the bias slowly. The derivative of this scaling factor in terms of $R^2$ is $\frac{\rho-1}{(1-R^2\rho)^2}$. To provide numerical intuition, if we take a case where $\rho=.9$ and $R^2$ is zero; the magnitude of the bias is the same as in the example with no matching. This is intuitive because if the matching variable has no explanatory power, it should not increase the accuracy of the treatment effect estimate. When $R^2=.5$ the bias is 90\% of the no matching bias, when $R^2=.7$ the bias is 81\% of the no matching bias, and when $R^2=.9$ the bias is still 53\% of the no matching bias. This suggests that matching is inefficient in reducing the local randomization bias.  
		 
\Sec{Re-Weighting the Sample by Reliability}

For parsimony, I used a classical test theory modal that combined all the measurement error into one term: $\omega_{i}$. However, measurement error has many components. There is measurement error associated with a specific occasion, like a barking dog. There is also measurement error associated with a specific set of test items, for example, a student getting unlucky and randomly getting none of the vocabulary words they know on the exam. Different methods of estimating test reliability will capture different facets of measurement error and will leave unmeasured ``hidden facets". Specifically, internal consistency reliability estimates capture item based measurement error, test-retest reliability estimates capture occasion based measurement error, and parallel forms reliability estimates capture both types of error \cite{haertel2006}.

\citet{eckles2020} solve the local randomization bias problem by developing a design based LATE estimator which uses estimates of the test reliability to re-weight the treatment and control samples to have equivalent latent variable distributions. However, most assessments, including the assessments from the Early Childhood Longitudinal Study that \citet{eckles2020} use in their study, only report internal consistency reliabilities. This is because internal consistency reliabilities can be estimated from the same assessment data used to generate an assessment proficiency estimate, unlike other types of reliability which require test takes retake the exam. Using these reliability estimates to re-weight the sample will therefore be problematic because they will be a downwardly biased estimate of the true measurement error. In fact, research has shown that occasion level error can be a larger source of error than the measured error from item sampling \cite{haertel2006}.

\Sec{Conclusion}

Substantial effort goes into making most test scores accurate measures of latent proficiency. This effort runs at cross-purposes to creating a randomized experiment. In RDDs analyzed using local linear regression, this is not a problem. However, when the heuristics used to explain the intuition behind RDD are taken literally, then the gap between heavily designed test scores and a truly randomized experiment can create problems.

In this paper, I quantify the bias associated with estimating the RDD LATE as a local random experiment. This quantification is not intended to adjust the LATE estimates ex-post. Such adjustment would move the estimate back into the model based inference and would not be superior to a typical LLR analysis. Instead, this paper is designed to clarify under what conditions a local randomization model is appropriate for estimating the RDD LATE. Test scores, which have high reliabilities and good predictive power, are unlikely to make for RDDs that can be analyzed with local randomization. Similarly, medical tests, which like academic proficiency tests, are designed to be precise and predictive, are unlikely to be good candidates for local randomization. On the other hand, RDDs, where the running variable is not a measure heavily designed by experts (e.g., birthday, election totals), may make for more plausible usages of the local randomization method.

\SmSec{Acknowledgements:}
The author would like to thank Eric Taylor, Andrew Ho, Luke Miratrix, and Nicole Pashley as well as the Miratrix
C.A.R.E.S. Lab for useful feedback. 

\SmSec{Funding information:} The author was supported by the American Educational Research Association which receives funds for its ``AERA-NSF Grants Program" from the National Science Foundation under NSF award NSF-DRL \#1749275 while working on this article. Opinions reflect those of the author and do not necessarily reflect those the American Educational Research Association or the National Science Foundation.

\SmSec{Conflict of interest:} The author states no conflict of interest

\newpage

\bibliographystyle{apacite}
\bibliography{Disbib}

\newpage

\begin{appendices}
\section{Derivation of Bias when the Relationship between the Outcome and Latent Proficiency Varies by Treatment Status}
	
	The assumption that $g_1(\theta_i)$ is constant and equals the treatment effect $\tau$ is dropped. Instead $g_1(\theta_i)$ is assumed to be a linear function $\beta+a_1\theta_i$. The new RDD model is written as follows:
	$$Y_i=a_0\theta_{i}+a_1\mathds{1}(X_i <c)\theta_{i}+\beta\mathds{1}(X_i <c) + \epsilon_i , \; \epsilon_{i} \stackrel{iid}{\sim}  N[0,\sigma_y^2]$$
  	$$X_i = \theta_i+ \omega_{i}, \; \omega_{i} \stackrel{iid}{\sim}  N[0,\sigma_{\omega}^2]$$

Under this model there is no longer a constant treatment effect and the LATE at $X=c-\Delta$ is therefore defined as:

	$$  \tau= E[Y_1 | X=c-\Delta] - E[Y_0 | X=c-\Delta]$$	
	$$  = E[a_0\theta+a_1\theta+\beta+ \epsilon_i  | X=c-\Delta] - E[a_0\theta+ \epsilon_i | X=c-\Delta]$$	
	$$  \tau = E[a_1\theta+\beta | X=c-\Delta]=a_1E[\theta | X=c-\Delta] + \beta$$

The LATE is estimated as: 

	$$  \hat{\tau}= E[Y_1 | X=c-\Delta] - E[Y_0 | X=c]$$
	$$=E[(a_0+a_1)\theta+\beta+\epsilon | X=c-\Delta]-E[a_0\theta+\epsilon | X=c]$$
	$$=(a_0+a_1)E[\theta | X=c-\Delta]+\beta-a_0E[\theta | X=c]$$

The estimator bias can be written as:
			$$  \hat{\tau}-\tau= a_0E[\theta | X=c-\Delta]-a_0E[\theta | X=c]+a_1E[\theta | X=c-\Delta]+\beta-a_1E[\theta | X=c-\Delta]-\beta$$
			$$ = a_0E[\theta | X=c-\Delta]-a_0E[\theta | X=c]$$
According to Kelley's formula \cite{kelley1947}: 	
	$$E[\theta | X=x] = \rho x +(1-\rho) E[X] \; , where \; \rho=\frac{\sigma_\theta^2}{\sigma_X^2} $$

Therefore the estimator bias can be rewritten as:	
	$$  \hat{\tau}-\tau=a_0(\rho(c-\Delta)+(1-\rho)E[X]-{\rho}c-(1-\rho)E[X])$$
	$$  \hat{\tau}-\tau=-a_0\rho\Delta$$

\label{sec:altderivation}
\end{appendices}
		
 \end{document}